\documentclass[aps,onecolumn,10pt]{revtex4}
\usepackage{amsmath}
\usepackage{amssymb}
\usepackage{hyperref}
\hypersetup{colorlinks=true} 
\usepackage{graphicx}
\usepackage{epsfig}
\usepackage{float}
\usepackage{slashed} 
\numberwithin{equation}{section}
\numberwithin{equation}{section}

\begin{document}
\allowdisplaybreaks
\setcounter{equation}{0}

\title{Ghost Problems from Pauli-Villars to Fourth-Order Quantum Gravity and their Resolution}

\author{Philip D. Mannheim \\}
\affiliation{Department of Physics, University of Connecticut, Storrs, CT 06269, USA \\
philip.mannheim@uconn.edu}
\smallskip
\date{March 31, 2020}

\begin{abstract}
We review the history of the ghost problem in quantum field theory from the Pauli-Villars regulator theory to currently popular fourth-order derivative quantum gravity theories.  While these theories all appear to have unitarity-violating ghost states with negative norm, we show that in fact these ghost states only appear because the theories are being formulated in the wrong Hilbert space.  In these theories the Hamiltonians are not Hermitian but instead possess an antilinear symmetry. Consequently, one cannot use inner products that are built out of states and their Hermitian  conjugates. Rather, one must use inner products built out of states and their conjugates with respect to the antilinear symmetry, and these latter inner products are positive. In this way one can build quantum theories of gravity in four spacetime dimensions that are unitary.
 \end{abstract}

\maketitle
\smallskip
\noindent
\centerline{Essay written for the Gravity Research Foundation 2020 Awards for Essays on Gravitation}

\bigskip

\section{The Seventy Year Ghost Problem}
\label{S1}

Some seventy years ago in the late 1940s Pauli and Villars \cite{Pauli1949} introduced a regularization scheme for Feynman diagrams that enabled one to maintain gauge invariance. While Pauli and Villars acknowledged in their paper that their regularization scheme was purely mathematical, they indicated that they did not want to rule out the possibility that it might actually be physical as well. In order to explore this possibility, in the early 1950s Pais and Uhlenbeck (PU) \cite{Pais1950} studied a fourth-order derivative theory that would yield the (schematic) Pauli-Villars propagator
\begin{eqnarray}
D(k)=\frac{1}{k^2-M_1^2}-\frac{1}{k^2-M_2^2},
\label{1.1}
\end{eqnarray} 
 only to discover that the associated Hamiltonian was unbounded from below. This was actually a generic finding since it had been known from the work of Ostrogradski in the 19th century \cite{Ostrogradski1850} that higher-derivative theories typically had such instabilities. At about the same time work was being developed by Lee \cite{Lee1954}, K\"all\'en and Pauli \cite{Kallen1955} and Heisenberg \cite{Heisenberg1957}, especially in regard to the Lee model \cite{Lee1954}, on the issue of ghost states. These ghost states are states with a Dirac norm (the overlap of a state with its Hermitian conjugate) that is negative, and would lead to a loss of unitarity and a loss of probability conservation.  And it was realized that one could in fact get the energy spectrum of the Pais-Uhlenbeck theory to be bounded from below if one were to quantize the theory with such negative norm states. The theory thus looked to be doomed since it suffered from one of two twin diseases, either negative norms or negative energies and it appeared that one could not get rid of both.

Motivated by the fact that a replacement of a $1/(k^2-M_1^2)$ propagator by (\ref{1.1}) would render logarithmic divergences in Feynman diagrams finite,  in the 1960s Lee and Wick \cite{Lee1969} reopened the issue of negative norm states, and found that the Pauli-Villars propagator could become unitary if the two particles that it embodied both became unstable and acquired masses that were in a complex conjugate pair. This of course left open the status of the theory if the two particle masses were to remain real, but since it did provide a solution to the unitarity problem, it has recently come back into prominence with the development of the Lee-Wick standard model \cite{Grinstein2008} of elementary particles. 

The solution to the problem in the two real masses case was found via a new approach to quantum theory that had been developed by  Bender and collaborators in the last twenty years especially as applied to the Lee model, namely via a possible role in quantum theory for non-Hermitian Hamiltonians that possess an antilinear symmetry such as $PT$ ($P$ is parity, $T$ is time reversal) \cite{Bender2007}. This approach was triggered by the surprising discovery that the energy eigenvalues of the non-Hermitian but $PT$-symmetric Hamiltonian $H=p^2+ix^3$ are all real \cite{Bender1998,Bender1999}. This approach can be formulated \cite{Mannheim2018a} on the basis on three key ingredients: (i) that Hermiticity of a Hamiltonian is only sufficient to yield real eigenvalues, with it being antilinearity that is the necessary condition, (ii) that specifying the ket state in a Schr\"odinger equation does not oblige the bra state to be its Hermitian conjugate, with its conjugate with respect to the antilinear symmetry instead leading to a time-independent, probability conserving inner product, and (iii) that commutation relations can be continued into the complex plane via complex similarity transformations and thus remain perfectly valid realizations of the quantum theory.

As $PT$ theory matured, Bender and coworkers \cite{Bender2005} realized that one could apply $PT$ techniques to the Lee model ghost problem (amazingly doing so no less than fifty years after its inception). As constructed, the Lee model enabled one to implement coupling constant renormalization in a closed form, so that one could relate the bare and dressed coupling constants analytically. Now the ghost that the Lee model possessed only occurred for certain values of the dressed coupling constant, and it was noted in \cite{Bender2005}  that for these values the bare coupling constant becomes complex, with the Hamiltonian  no longer being Hermitian. However, it turns out that the complex bare coupling constant phase is $PT$ symmetric, and when one uses the $PT$ theory norm (the overlap of a ket with its $PT$ conjugate) the inner product is then positive and there are no ghost states at all. The ability of $PT$ theory to solve the Lee model ghost problem  is a considerable triumph for it, indicating that if one were to obtain a negative Dirac norm in a calculation that would not necessarily mean that the theory was not unitary. Rather it could mean that one is in the wrong Hilbert space, and that one could be fully unitary in a Hilbert space with inner products that are based on kets and their $PT$ (or some other appropriate antilinear symmetry operator) conjugates. Thus if a non-Hermitian Hamiltonian has an antilinear symmetry such as $PT$ one should not use as inner product the overlap of a ket with its Hermitian conjugate. Rather, one should use the overlap of ket with its antilinear symmetry conjugate. In such a case one still has a fully viable quantum theory. Since Hermitian conjugation is itself an antilinear procedure, Hermiticity becomes a special case of antilinear symmetry, with Hamiltonians being able to be both Hermitian and have an antilinear symmetry. However, antilinear symmetry is the more general as it allows for Hamiltonians to not be Hermitian and still be associated with a viable, probability conserving, quantum theory.

While there would appear to be a wide variety of antilinear symmetry operators that one could consider (and even when does not use $PT$ itself such theories are generically referred to as $PT$ theories), it turns out that $CPT$ symmetry ($C$ is charge conjugation) is uniquely selected by two very general requirements, namely invariance under the complex Lorentz group (the linear part of a $CPT$ transformation being a particular complex Lorentz transformation) and conservation of probability \cite {Mannheim2018a}. With there being no need to impose any Hermiticity requirement the $CPT$ theorem is thus extended to non-Hermitian Hamiltonians (the original proofs of the $CPT$ theorem from the 1950s assumed Hermiticity). And in those cases in which $C$ is separately conserved (the cases typically studied by Bender and coworkers), $CPT$ defaults to $PT$, with $PT$ symmetry thus being put on a quite firm theoretical footing.

The success that $PT$ theory had in dealing with the Lee model ghost problem immediately raises the question of whether the Pauli-Villars ghost problem itself could be solved the same way. And as shown by Bender and Mannheim \cite{Bender2008a,Bender2008b}  this is in fact the case. Specifically, they found that the Hamiltonian of the Pais-Uhlenbeck model is not Hermitian but is instead $PT$ symmetric. The wave functions of the theory were found to not be normalizable on the real coordinate axis, and in such a basis one could not integrate by parts with the Hamiltonian then not being Hermitian in this basis. Thus despite the fact that all coefficients in the Hamiltonian are real, asymptotic boundary condition behavior could still prevent the Hamiltonian from being Hermitian. However, the states would become normalizable if the operators of the theory were to be continued into a specific domain in the complex plane (known as a Stokes wedge domain), and such a continuation is perfectly permissible in quantum theory since commutation relations are preserved under complex similarity transformations, and can thus be perfectly viable realizations of the quantum theory. In such a Stokes wedge domain one now can integrate by parts, and with energy eigenfunctions now being well-defined, one finds that the energy spectrum is bounded from below (no Ostrogradski instability). With the $PT$ norm being positive definite, both of the twin diseases of the Pais-Uhlenbeck theory (negative energies or negative norms) are thus solved simultaneously, and the theory is viable.

While $PT$ theory encompasses Hermitian quantum theory as a special case since $PT$-symmetric Hamiltonians can also be Hermitian ($PT$ theory is not in any way a modification of conventional quantum mechanics, it just takes advantage of the freedom that is present in its Hilbert space formulation as the bra is not obliged to be the Hermitian conjugate of the ket), it nonetheless admits of realizations that cannot be achieved in the Hermitian case. Specifically, one can have  all energies real and energy eigenspectrum complete, energies in complex conjugate pairs and energy eigenspectrum still complete, or all energies real but energy eigenspectrum incomplete (the non-diagonalizable Jordan-block case). The latter two cases are not achievable with a Hermitian Hamiltonian. The Jordan-block realization is relevant to the fourth-order derivative conformal gravity theory that is discussed below. And as to the Lee-Wick model, in its realization in which the energies are in a complex conjugate pair the model is a $PT$ theory. Since the work of \cite{Bender2008a,Bender2008b} shows that even in the real mass case the theory is a $PT$ theory, the Pauli-Villars propagator emerges as a $PT$-theory propagator no matter whether masses are real or in complex conjugate pairs. 

Beyond the issue of the basic structure of the Pauli-Villars propagator itself, there is also the question of whether unitarity is preserved by  radiative loop corrections that involve it. However, since one cannot change the signature of an inner product perturbatively, unitarity cannot be lost. Nonetheless, the mechanism for actually achieving unitarity is somewhat surprising, since on their own loop diagrams are not unitarity (the discontinuity across the propagator of (\ref{1.1}) is not positive definite). However, there is unexpected and novel contribution from the tree approximation graph (a contribution that is foreign to Hermitian theories), and when tree and loop graphs are taken together, unitarity is maintained  \cite{Mannheim2018b}. Theories based on the Pauli-Villars propagator can thus be regarded as being fully viable.

Another area where the $D(k)$ propagator given in (\ref{1.1}) appears is in  gravity theories whose actions involve not just the Ricci scalar (standard Einstein gravity) but also quadratic powers of it or quadratic powers of the Riemann or Ricci tensors. With the standard $1/k^2$ propagator of non-renormalizable Einstein gravity leading to uncontrollable quadratic divergences in Feynman diagrams, its replacement by (\ref{1.1}) would cause quadratic divergences in Feynman diagrams to only be logarithmically divergent and thus render them renormalizable. However then, by being based on (\ref{1.1}) these theories equally have a potential ghost problem \cite{Stelle1977}, and this problem is also resolved by reinterpreting these theories as $PT$ theories  (see also \cite{Hawking2002}). Such fourth-order derivative theories of gravity are of interest because they can have an underlying scale symmetry, to thus give them dimensionless coupling constants that make them power-counting renormalizable. One particularly interesting case is conformal gravity, a theory with local (i.e. not just global) scale invariance (see the recent reviews in \cite{Mannheim2006,Mannheim2012,Mannheim2017}), and this theory is actually a Jordan-block theory \cite{Mannheim2005}, \cite{Mannheim2006}, \cite{Bender2008b} and thus completely outside of the Hermitian framework, but part and parcel of the $PT$, and thus its unitary and ghost free, framework. Recently there has been a burst of interest in gravity theories that are globally scale invariant (there has recently even been an entire conference at CERN dedicated to this issue \cite{Cern2019}), and ghost issues in all of these theories can be resolved by the $PT$ approach. $PT$ symmetry is thus very rich and well worth further study. With quantum conformal gravity always having been renormalizable (its coupling constant being dimensionless) establishing that it is in fact unitary (i.e. ghost free) as well opens the door \cite{Mannheim2012,Mannheim2017} to the construction of a fully consistent and unitary theory of quantum gravity in four spacetime dimensions.

\section{The Pauli-Villars Regulator Scheme and the Pais-Uhlenbeck Fourth-Order Oscillator}
\label{S2}

In trying to regulate the asymptotic momentum behavior of  Feynman diagrams while being able to maintain gauge invariance in the gauge theory case, Pauli and Villars \cite{Pauli1949} suggested that one replace the generic scalar field $D(k)=1/(k^2-M_1^2)$ propagator by the propagator given in (\ref{1.1}), together with analogous expressions for fermions and gauge bosons. As conceived by Pauli and Villars it was necessary that both of the $1/(k^2-M_1^2)$ and $1/(k^2-M_2^2)$ propagators couple to vertices with the same relative sign. The two propagators would act as mirror images of each other, to thus be associated with two independent and decoupled second-order derivative actions
\begin{eqnarray}
I_{S_1}+I_{S_2}=\int d^4x\left[\tfrac{1}{2}\partial_{\mu}\phi_1\partial^{\mu}\phi_1-\tfrac{1}{2}M_1^2\phi_1^2-\lambda\phi_1^4\right]+\int d^4x\left[\tfrac{1}{2}\partial_{\mu}\phi_2\partial^{\mu}\phi_2-\tfrac{1}{2}M_2^2\phi_2^2-\lambda\phi_2^4\right].
\label{2.1}
\end{eqnarray}
With the propagator associated with (\ref{2.1}) being required to be given by 
\begin{eqnarray}
D(k)=\frac{1}{k^2-M_1^2}-\frac{1}{k^2-M_2^2},
\label{2.2}
\end{eqnarray} 
it would lead to the desired cancellation of infinities in Feynman diagrams as it would behave asymptotically as $1/k^4$ rather than as $1/k^2$. With the propagator being associated with the Fourier transform of $\langle\Omega_1|T[\phi_1(x)\phi_1(0)]|\Omega_1\rangle$+$\langle\Omega_2|T[\phi_2(x)\phi_2(0)]|\Omega_2\rangle$, and with the residues of the poles in $-1/(k^2-M_2^2)$ being negative, the closure relation for states of the system would be of the form
\begin{eqnarray}
\sum |n_1\rangle\langle n_1|-\sum |n_2\rangle\langle n_2|=I,
\label{2.3}
\end{eqnarray}
and while one could take $\phi_1$ to be quantized with positive norm one would however have to take $\phi_2$ to be quantized in a negative norm Krein space. Thus if one uses the action $I_{S_2}+I_{S_1}$ the relative minus sign in $D(k)$ would be associated with negative norm states and would violate unitarity.

The objective of Pais and Uhlenbeck was to see whether one could generate the same $D(k)=1/(k^2-M_2^2)-1/(k^2-M_1^2)$ propagator from an action involving a single neutral field $\phi(x)$. With the action
\begin{eqnarray}
I_S&=&\tfrac{1}{2}\int d^4x\bigg{[}\partial_{\mu}\partial_{\nu}\phi\partial^{\mu}
\partial^{\nu}\phi-(M_1^2+M_2^2)\partial_{\mu}\phi\partial^{\mu}\phi
+M_1^2M_2^2\phi^2\bigg{]},
\label{2.4}
\end{eqnarray}
having a fourth-order derivative equation of motion given by
\begin{eqnarray}
&&(\partial_t^2-\vec{\nabla}^2+M_1^2)(\partial_t^2-\vec{\nabla}^2+M_2^2)
\phi(x)=0,
\label{2.5}
\end{eqnarray}
the associated propagator is given by
\begin{eqnarray}
&&D(k)=\frac{1}{(k^2-M_1^2)(k^2-M_2^2)}=\frac{1}{(M_1^2-M_2^2)}\left(\frac{1}{k^2-M_1^2}-\frac{1}{k^2-M_2^2}\right),
\label{2.6}
\end{eqnarray}
with (\ref{2.6}) thus being recognized as being of the same form as (\ref{2.2}). However, since there is now only one field that is involved, one would (incorrectly as we shall see) identify (\ref{2.6}) with the Fourier transform of the one-field $\langle\Omega|T[\phi(x)\phi(0)]|\Omega\rangle$ rather than with that of the two-field $\langle\Omega_1|T[\phi_1(x)\phi_1(0)]|\Omega_1\rangle$+$\langle\Omega_2|T[\phi_2(x)\phi_2(0)]|\Omega_2\rangle$. With $\phi$ obeying a fourth-order derivative equation of motion it has two sets of eigenstates. And if they are labelled $|n_1\rangle$ and $|n_2\rangle$,  taking them to obey (\ref{2.3}) would lead to (\ref{2.6}) on the insertion of (\ref{2.3}) into $\langle\Omega|T[\phi(x)\phi(0)]|\Omega\rangle$. 

Since only time derivatives are relevant to quantization and not spatial derivatives,  on setting $\omega_1=(\bar{k}^2+M_1^2)^{1/2}$, $\omega_2=(\bar{k}^2+M_2^2)^{1/2}$ and dropping the spatial dependence, the $I_S$ action reduces to the acceleration-dependent quantum-mechanical Pais-Uhlenbeck two-oscillator model action \cite{Pais1950} 
\begin{eqnarray}
I_{\rm PU}=\tfrac{1}{2}\int dt\left[{\ddot z}^2-\left(\omega_1^2
+\omega_2^2\right){\dot z}^2+\omega_1^2\omega_2^2z^2\right],
\label{2.7}
\end{eqnarray}
where for the moment we take $\omega_1$ and $\omega_2$ to be real with $\omega_1>\omega_2$. The equation of motion is given by
\begin{eqnarray}
\overset{....}{z}+(\omega_1^2+\omega_2^2)\overset{..}{z}+\omega_1^2\omega_2^2
z=0,
\label{2.8}
\end{eqnarray}
while in analog to (\ref{2.2}) and (\ref{2.6}) the propagator is given by
\begin{eqnarray}
&&
G(E)=\frac{1}{(E^2-\omega_1^2)(E^2-\omega_2^2)}=\frac{1}{(\omega_1^2-\omega_2^2)}\left(\frac{1}{E^2-\omega_1^2}-\frac{1}{E^2-
\omega_2^2}\right),
\label{2.9}
\end{eqnarray}
and is just as  problematic as (\ref{2.6}). (In (\ref{2.7}) it is understood that the operator $z$ is just a stand-in for $\phi$, with the continuation into the complex plane that we make for it below actually being a continuation of $\phi$ and not of the spacetime coordinates on which $\phi$ depends.) 

As given, $I_{\rm PU}$ is a constrained system since $\dot{z}$ would have to serve as the conjugate of both $z$ and $\ddot{z}$, i.e. $I_{\rm PU}$ has too many degrees of freedom for one oscillator but not enough for two. To  construct a Hamiltonian $H_{\rm PU}$ for this system one has to use the method of Dirac constraints. And on setting $x=\dot{z}$ this leads to  \cite{Mannheim2005}
\begin{eqnarray}
H_{\rm PU}=\tfrac{1}{2}p_x^2+p_zx+\tfrac{1}{2}\left(\omega_1^2+\omega_2^2 \right)x^2-\tfrac{1}{2}\omega_1^2\omega_2^2z^2, \quad [z,p_z]=i, \quad [x,p_x]=i.
\label{2.10}
\end{eqnarray}
We note that with all the poles in the above $G(E)$ being on the real $E$ axis, all the eigenvalues of $H_{\rm PU}$ are real.

If we now make the standard substitutions
\begin{eqnarray}
z&=&a_1+a_1^{\dagger}+a_2+a_2^{\dagger}\quad p_z=i\omega_1\omega_2^2
(a_1-a_1^{\dagger})+i\omega_1^2\omega_2(a_2-a_2^{\dagger}),
\nonumber\\
x&=&-i\omega_1(a_1-a_1^{\dagger})-i\omega_2(a_2-a_2^{\dagger}),\quad
p_x=-\omega_1^2 (a_1+a_1^{\dagger})-\omega_2^2(a_2+a_2^{\dagger}),
\label{2.11}
\end{eqnarray}
we obtain a Hamiltonian and commutator algebra \cite{Mannheim2005} 
\begin{align}
H_{\rm PU}&=2(\omega_1^2-\omega_2^2)(\omega_1^2 a_1^{\dagger}
a_1-\omega_2^2a_2^{\dagger} a_2)+\tfrac{1}{2}(\omega_1+\omega_2),
\nonumber\\
[a_1,a_1^{\dagger}]&=\frac{1}{2\omega_1(\omega_1^2-\omega_2^2)},\quad [a_2,a_2^{\dagger}]=-\frac{1}{2\omega_2(\omega_1^2-\omega_2^2)},
\label{2.12}
\end{align}
and note that the $[a_2,a_2^{\dagger}]$ commutator is negative. 

If we take $a_1$ and $a_2$ to annihilate the no-particle state $|\Omega\rangle$ according to $a_1|\Omega\rangle=0$, $a_2|\Omega\rangle=0$, the energy spectrum that ensues is then bounded from below with $|\Omega\rangle$ being the ground state with energy $\left(\omega_1+\omega_2\right)/2$. However, the excited state
$a_2^\dag|\Omega\rangle$, which lies at energy $\omega_2$ above the ground state, has a Dirac norm $\langle\Omega|a_2a_2^\dag|\Omega\rangle$ that is negative. 

Alternatively, if we take $a_1$ and $a_2^{\dagger}$ to annihilate the no-particle state $|\Omega\rangle$, according to $a_1|\Omega\rangle=0$, $a_2^\dag|\Omega\rangle=0$,  the theory would then be free of negative-norm
states, but the energy spectrum would be unbounded below (the Ostrogradski instability problem). As noted above, the theory thus suffers from one of two twin diseases, either negative norms or negative energies and it appears that one could not get rid of both. Since defining the vacuum by setting $a_2|\Omega\rangle=0$ or by setting $a_2^{\dagger}|\Omega\rangle=0$ would correspond to working in two totally different Hilbert spaces, in no single Hilbert space does one have both problems, though in either one there is still a seemingly irrefutable  problem.

However, as noted in \cite{Bender2008a,Bender2008b} this seemingly irrefutable analysis actually has a flaw. Specifically, if we now set $p_z=-i\partial_z$, $p_x=-i\partial_x$, the Schr\"odinger equation takes the form
\begin{eqnarray}
\left[-\frac{1}{2}\frac{\partial^2}{\partial x^2}-ix\frac{\partial}{
\partial z}+\frac{1}{2}(\omega_1^2+\omega_2^2)x^2-\frac{1}{2}
\omega_1^2\omega_2^2z^2\right]\psi_n(z,x)=E_n\psi_n(z,x),
\label{2.13}
\end{eqnarray}
with the lowest positive energy state with $E_0=(\omega_1+\omega_2)/2$ having eigenfunction \cite{Bender2008a} 
\begin{eqnarray}
\psi_0(z,x)={\rm exp}\left[\frac{1}{2}(\omega_1+\omega_2)\omega_1\omega_2
z^2+i\omega_1\omega_2zx-\frac{1}{2}(\omega_1+\omega_2)x^2\right].
\label{2.14}
\end{eqnarray}
As $z\to\pm\infty$, $\psi_0(z,x)$ diverges, with, as noted earlier,  the wave function of the ground state $|\Omega\rangle$ (and thus its $\langle \Omega |\Omega\rangle=\int dxdz\langle\Omega |xz\rangle\langle xz|\Omega\rangle=\int dxdz\psi^*_0(x,z)\psi_0(x,z)$ Dirac norm)  not being normalizable on the real $z$ axis. Such lack of normalizability means that the closure relation given in (\ref{2.3}) could not hold as it presupposes normalizable states. As we now show, it is this lack of normalizability that actually saves the theory. 

To make the states be normalizable one must continue $z$ into the complex plane. With $z=re^{i\theta}$, $\psi_0(z,x)$ will be normalizable if $\cos 2\theta <0$, i.e. if $\theta$ lies in wedges with $\pi/4 <\theta < 3\pi/4$,  $5\pi/4 <\theta < 7\pi/4$ (shaped like the top and bottom quadrants of the letter $X$), i.e. wedges that include the imaginary $z$ axis but not the real $z$ axis. In these wedges one now can integrate by parts. Now we note that because of the $-\tfrac{1}{2}\omega_1^2\omega_1^2z^2$ term as given in (\ref{2.10}),  $H_{PU}$ would be unbounded from below with real $z$, but not if $z$ is pure imaginary. Thus in the $\pi/4 <\theta < 3\pi/4$,  $5\pi/4 <\theta < 7\pi/4$ wedges there is no Ostrogradski instability. Now the reader might object that an operator such as $z$ is Hermitian. However, while it would be in the basis of its own eigenstates (which are on the real $z$ axis), that does not make it Hermitian (i.e. integrable by parts) when acting on the eigenstates of $H_{PU}$ instead when they are also written on the same real $z$ axis. In fact it is the mismatch between the basis in which the individual components of a Hamiltonian are Hermitian and the basis of the eigenstates of the Hamiltonian itself that is the hallmark of $PT$ studies.  

To implement the continuation we  make a similarity transform on the operators in $H_{PU}$ of the form \cite{Bender2008a,Bender2008b}
\begin{eqnarray}
y=e^{\pi p_zz/2}ze^{-\pi p_zz/2}=-iz,\qquad q=e^{\pi p_zz/2}p_ze^{-\pi p_zz/2}=
ip_z,
\label{2.15}
\end{eqnarray}
so that $[y,q]= i$. Under this same transformation $H_{\rm PU}$ transforms into
\begin{eqnarray}
e^{\pi p_zz/2}H_{\rm PU}e^{-\pi p_zz/2}=\bar{H}=\frac{p^2}{2}-iqx+\frac{1}{2}\left(\omega_1^2+\omega_2^2
\right)x^2+\frac{1}{2}\omega_1^2\omega_2^2y^2,
\label{2.16}
\end{eqnarray}
where for notational simplicity we have replaced $p_x$ by $p$, so that $[x,p]=i$. When acting on the eigenfunctions of $\bar{H}$ the  $y$ and $q$ operators are Hermitian (as are $x$ and $p$). However, as the presence of the factor $i$ in the $-iqx$ term indicates, $\bar{H}$ is not Hermitian even though all of its eigenvalues are real (the similarity transformation in (\ref{2.16}) is isospectral). While not being Hermitian $\bar{H}$ is $PT$ symmetric, with $x$ and $y$ being $PT$ odd and $p$ and $q$ being $PT$ even \cite{Bender2008a,Bender2008b}, to thus provide a straightforward example of a non-Hermitian but $PT$-symmetric Hamiltonian whose eigenvalues are all real. 

To  quantize the theory one sets \cite{Bender2008b}
\begin{eqnarray}
\dot{y}(t)&=&i[\bar{H},y]=-ix(t),\qquad \dot{x}(t)=p(t),\qquad
\dot{p}(t)=iq(t)-(\omega_1^2+\omega_2^2)x(t),\qquad
\dot{q}(t)=-\omega_1^2\omega_2^2y(t),
\nonumber\\
y(t)&=&-ia_1e^{-i\omega_1t}+a_2e^{-i\omega_2t}-i\hat{a}_1e^{i\omega_1t}+\hat{a}_2e^{i\omega_2t},
\nonumber\\
x(t)&=&-i\omega_1a_1e^{-i\omega_1t}+\omega_2a_2e^{-i\omega_2t}+i\omega_1\hat{a}_1
e^{i\omega_1t}-\omega_2\hat{a}_2e^{i\omega_2t},
\nonumber\\
p(t)&=&-\omega_1^2a_1e^{-i\omega_1t}-i\omega_2^2a_2e^{-i\omega_2t}-\omega_1
^2\hat{a}_1e^{i\omega_1t}-i\omega_2^2\hat{a}_2e^{i\omega_2t},
\nonumber\\
q(t)&=&\omega_1\omega_2[-\omega_2a_1e^{-i\omega_1t}-i\omega_1a_2e^{-i
\omega_2t}+\omega_2\hat{a}_1e^{i\omega_1t}+i\omega_1\hat{a}_2e^{i\omega_2t}].
\label{2.17}
\end{eqnarray}
With $[x,p]=i$, $[y,q]=i$, the $a_i$ and $\hat{a}_i$ operators obey the standard two-oscillator commutation algebra 
\begin{eqnarray}
&&[a_1,\hat{a}_1]=\frac{1}{2\omega_1(\omega_1^2-\omega_2^2)},\quad
[a_2,\hat{a}_2]=\frac{1}{2\omega_2(\omega_1^2-\omega_2^2)},
\nonumber\\
&&[a_1,a_2]=0,\qquad [a_1,\hat{a}_2]=0,\qquad [\hat{a}_1,a_2]=0,\quad [\hat{a}_1,\hat{a}_2]=0,
\label{2.18}
\end{eqnarray}
where now there are no minus signs in commutators. Similarly, the Hamiltonian takes the form 
\begin{eqnarray}
\bar{H}=2(\omega_1^2-\omega_2^2)[\omega_1^2\hat{a}_1a_1+\omega_2^2\hat{a}_2a_2]
+\tfrac{1}{2}(\omega_1+\omega_2),
\label{2.19}
\end{eqnarray}
and now all energy eigenvalues are positive ($\omega_1>\omega_2$). We thus obtain a Hamiltonian with no states of negative norm and no states of negative energy, with the theory this being fully quantum-mechanically viable.

To underscore that all norms are positive we note that we can make a similarity transformation on $\bar{H}$ in order to decouple the two oscillators \cite{Bender2008a,Bender2008b}. Specifically, one introduces an operator $Q$
\begin{eqnarray}
Q=\alpha pq+\beta xy,\qquad \alpha=\frac{1}{\omega_1\omega_2}{\rm log}\left(\frac{\omega_1+\omega_2}{\omega_1-\omega_2}\right),\qquad \beta=\alpha\omega_1^2\omega_2^2,
\label{2.20}
\end{eqnarray}
with $Q$ being Hermitian since $x$, $y$, $p$ and $q$  are all Hermitian, while being $PT$ even. With this $Q$ $\bar{H}$  transforms to \cite{Bender2008a} 
\begin{eqnarray}
e^{-Q/2}\bar{H}e^{Q/2}&=&\bar{H}^{\prime}
=\frac{p^2}{2}+\frac{q^2}{2\omega_1^2}+
\frac{1}{2}\omega_1^2x^2+\frac{1}{2}\omega_1^2\omega_2^2y^2.
\label{2.21}
\end{eqnarray}
We recognize $\bar{H}^{\prime}$ as being a fully acceptable standard two-dimensional oscillator system. However, we cannot just treat the system as a two decoupled oscillators since under the same transformation an interaction term such as $\lambda y^4$ would transform into $\lambda [y^{\prime}]^4$ where 
\begin{eqnarray}
y^{\prime}&=&e^{-Q/2}ye^{Q/2}=y\cosh\theta+i(\alpha/\beta)^{1/2}p\sinh\theta, \quad \theta=\tfrac{1}{2}(\alpha\beta)^{1/2}.
\label{2.22}
\end{eqnarray}
However there can be no loss of unitarity under the radiative corrections associated with $\lambda[y^{\prime}]^4$ as one cannot change the signature of a Hilbert space in perturbation theory. How this is achieved in practice is described in \cite{Mannheim2018b}.

In addition we note that with its phase being $-Q/2$ rather than $-iQ/2$, the $e^{-Q/2}$ operator is not unitary. The transformation from $\bar{H}$ to $\bar{H}^{\prime}$ is thus not a unitary transformation, but is a transformation from a skew basis with eigenvectors $|n\rangle$ to an orthogonal basis with eigenvectors $|n^{\prime}\rangle=e^{-Q/2}|n\rangle$, $\langle n^{\prime}|=\langle n|e^{-Q/2}$. Then since $\langle n^{\prime}|m^{\prime}\rangle =\delta_{mn}$, the eigenstates of $\bar{H}$ obey
\begin{eqnarray}
\langle n|e^{-Q}|m\rangle=\delta_{mn},\quad \sum_n|n\rangle\langle n|e^{-Q}=I,\quad \bar{H}=\sum _n|n\rangle E_n\langle n|e^{-Q}, \quad
\bar{H}|n\rangle=E_n|n\rangle,\quad \langle n|e^{-Q}\bar{H}=\langle n|e^{-Q}E_n.
\label{2.23}
\end{eqnarray}
We thus recognize the inner product as being not $\langle n|m\rangle$ but $\langle n|e^{-Q}|m\rangle$, with the conjugate of $|n\rangle$ being $\langle n|e^{-Q}$. This state is also the $PT$ conjugate of $|n\rangle$, so that the inner product is the overlap of a state with its $PT$ conjugate just as we had noted earlier. And as such this inner product is positive definite since $\langle n^{\prime}|m^{\prime}\rangle =\delta_{mn}$ is. 

Given (\ref{2.23}) we see that the propagator given in (\ref{2.9}) is not in fact the Fourier transform of $\langle\Omega |T[y(t)y(0)]|\Omega\rangle$. Rather it is the transform of $\langle\Omega |e^{-Q}T[y(t))y(0)]|\Omega\rangle$, with it being the presence of the $e^{-Q}$ factor that generates the minus sign in (\ref{2.9})  and not the presence of states with negative norm \cite{Bender2008b}. Similarly, as discussed in detail in \cite{Mannheim2018b}, for the $I_S$ action given in (\ref{2.4}) the propagator given in (\ref{2.6}) is identified not with $\langle\Omega|T[\phi(x)\phi(0)]|\Omega\rangle$ but with $\langle\Omega |VT[\phi(x)\phi(0)]|\Omega\rangle$  as evaluated with the $V$ that implements $VHV^{-1}=H^{\dagger}$ for the $H$ associated with $I_{S}$ (see the discussion in \cite{Mannheim2018a} and below), with the transformation for the Pais-Uhlenbeck $\bar{H}$ being
\begin{eqnarray}
e^{-Q}\bar{H}e^{Q}&=&\frac{p^2}{2}+iqx+\frac{1}{2}\left(\omega_1^2+\omega_2^2
\right)x^2+\frac{1}{2}\omega_1^2\omega_2^2y^2=
\bar{H}^{\dagger}.
\label{2.24}
\end{eqnarray}

Thus the resolution of the ghost problem is that the $D(k)$  propagator had been incorrectly represented as $\langle\Omega|T[\phi(x)\phi(0)]|\Omega\rangle$. Now since $D(k)$ is a c-number one can only identify it as the vacuum matrix element of a q-number product of fields after first constructing the underlying quantum Hilbert space, and not the other way round.  And when one does construct the Hilbert space one finds that in fact $D(k)$ has to be identified with $\langle\Omega |VT[\phi(x)\phi(0)]|\Omega\rangle$ instead. And then there is no ghost problem. Now while the $D(k)$ propagator can also be associated with the second-order derivative two-field action $I_{S_1}+I_{S_2}$ given in (\ref{2.1}), in that case there would be states of negative norm. However, when associated with the single fourth-order derivative single-field action $I_S$ given in (\ref{2.4}) there are no negative norm states. Thus the fact that both $I_{S_1}+I_{S_2}$ and $I_S$ lead to the same propagator ((\ref{2.2}) and (\ref{2.6})) does not mean that they therefore describe the same theory or that ghosts in one implies ghosts in the other.

\section{Antilinear Symmetry and the Lee-Wick Model}
\label{S3}

To see how antilinearity works in general it is instructive to look at the eigenvector equation 
\begin{eqnarray}
i\frac{\partial}{\partial t}|\psi(t)\rangle=H|\psi(t)\rangle=E|\psi(t)\rangle.
\label{3.1}
\end{eqnarray}
On replacing the parameter $t$ by $-t$ and then multiplying by a general antilinear operator $A$ we obtain
\begin{eqnarray}
i\frac{\partial}{\partial t}A|\psi(-t)\rangle=AHA^{-1}A|\psi(-t)\rangle=E^*A|\psi(-t)\rangle.
\label{3.2}
\end{eqnarray}
From (\ref{3.2}) we see that if $H$ has an antilinear symmetry so that $AHA^{-1}=H$, then, as first noted by Wigner in his study of time reversal invariance, energies can either be real and have eigenfunctions that obey $A|\psi(-t)\rangle=|\psi(t)\rangle$, or can appear in complex conjugate pairs that have conjugate eigenfunctions ($|\psi(t)\rangle \sim \exp(-iEt)$ and $A|\psi(-t)\rangle\sim \exp(-iE^*t)$). As noted in \cite{Mannheim2018a} the necessary and sufficient condition for all eigenvalues to be real is that $H$ has an antilinear symmetry and that its eigenstates are also eigenstates of the antilinear operator. Now if $H$ and $H^{\dagger}$ are related by a similarity transformation according to $VHV^{-1}=H^{\dagger}$ then all the eigenvalues of $H$ are either real or in complex conjugate pairs too, i.e. precisely the same outcome as when $H$ has an antilinear symmetry. Thus, as noted in \cite{Mannheim2018a}, the conditions $VHV^{-1}=H^{\dagger}$ and $AHA^{-1}=H$ are equivalent, in consequence of which the Pais-Uhlenbeck oscillator is both $PT$ symmetric and obeys (\ref{2.24}). We now discuss the relevance of these remarks to the Lee-Wick model.

Rather than work with the  $D(k)=1/(k^2-M_1^2)-1/(k^2-M_2^2)$ propagator,  Lee and Wick complexified the masses into a complex conjugate pair according to 
\begin{eqnarray}
\tilde{D}(k)=\frac{1}{k^2-M^2-iN^2}-\frac{1}{k^2-M^2+iN^2},
\label{3.3}
\end{eqnarray}
and were then able to establish unitarity. We now see that with this complexification the poles in $\tilde{D}(k)$ are in complex conjugate pairs, and thus the Lee-Wick theory has an antilinear symmetry. It is this symmetry that causes the theory to be unitary \cite{Mannheim2018a}. However since the $D(k)=1/(k^2-M_1^2)-1/(k^2-M_2^2)$ propagator will also lead to a unitary theory if it is associated with the $I_S$ action and not the $I_{S_1}+I_{S_2}$ one, even with real masses the theory is already unitary and complexification is not necessary. It is thus interesting to note that both the Lee model and the Lee-Wick model have an underlying antilinear symmetry, and in both cases that protects unitarity.

To underscore the point we note that if we set $\omega_1=\alpha+i\beta$, $\omega_2=\alpha-i\beta$ we can bring the $G(E)$ propagator given in (\ref{2.9}) to the (\ref{3.3}) form. However, when we do so we obtain $\omega_1^2+\omega_2^2=2(\alpha^2-\beta^2)$, $\omega_1^2\omega_2^2=(\alpha^2+\beta^2)^2$. In consequence, these factors remain real in $\bar{H}$ as given in (\ref{2.16}), and the theory remains $PT$ symmetric. Moreover with $\omega_1+\omega_2=2\alpha$ and $\omega_1\omega_2=\alpha^2+\beta^2$, the asymptotic behavior of the $\psi_0(x,z)$ wave function given in (\ref{2.14}) remains unchanged and a continuation into the same Stokes wedges as in the real $\omega_1$, $\omega_2$ case goes through without modification.

Also in regard to the Lee-Wick mechanism, we recall that in the literature there have been some concerns expressed as to whether or not it is causal. To this end we note that with real masses both of the $1/(k^2-M_1^2)$ and $1/(k^2-M_2^2)$ propagators are separately causal if one uses the Feynman contour for each one, and the pole and thus causality structure does not change if one replaces $1/(k^2-M_2^2)$ by  $-1/(k^2-M_2^2)$. However, if one were to work with the $\tilde{D}(k)=1/(k^2-M^2-iN^2)-1/(k^2-M^2+iN^2)$ propagator, then, as noted in \cite{Mannheim2013}, to maintain causality the Feynman contour would have to be deformed so that the upper-right-quadrant complex $k_0$ plane poles at $k^0=(\textbf{k}^2+M^2+iN^2)^{1/2}$ would be enclosed in a contour integration that closes in the lower-half complex $k^0$ plane (i.e. the same number of poles inside the contour as in the real mass case), and causality can then be secured.

\section{The Jordan-Block Case}
\label{S4}

In the analysis of the implications of antilinear symmetry in (\ref{3.2}) we had found two particular realizations, all energies real or energies in complex conjugate pairs. So if we vary parameters (such as $\omega_1$ and $\omega_2$ as described above) we can transit from one realization to the other. So we need to determine what happens at the transition point (a point known as an exceptional point in the $PT$ literature). For any given pair of complex conjugate energies the energies must become equal at the transition point, and with the eigenfunctions being complex conjugates of each other, they must become equal too. Thus at the transition point we lose an eigenfunction and the Hamiltonian becomes of non-diagonalizable, Jordan-block form. That it could not be diagonalized can be seen from the structure of $Q$ as given by (\ref{2.20}) in the  Pais-Uhlenbeck case, as $e^{-Q}$ becomes singular when the energies become equal, i.e. when $\omega_1=\omega_2=\omega$. Moreover, as we see from the structure of $G(E)$ given in (\ref{2.9}) the partial fraction decomposition becomes undefined at $\omega_1^2=\omega_2^2$ with the propagator becoming $1/(E^2-\omega^2)^2$. This is equally true of the relativistic (\ref{2.6}). Thus the decomposition into two sets of poles and the wisdom inferred from such a decomposition is not valid at the transition point. Since we lose an eigenstate at the transition point we have to ask where the other eigenfunction went to. Thus if we take the generic wave functions $e^{i(\omega+\epsilon)t}$ and  $e^{i(\omega-\epsilon)t}$, in the limit $\epsilon \rightarrow 0$ both wave functions reduce to $e^{i\omega t}$. To get a second solution that does not vanish in the $\epsilon \rightarrow 0$ we take
\begin{eqnarray}
\frac{e^{i(\omega+\epsilon)t}-e^{i(\omega-\epsilon)t}}{\epsilon}\rightarrow 2it e^{i\omega t}.
\label{4.1}
\end{eqnarray}
Because of the multiplicative $t$ factor this function is not stationary, and is hence not an energy eigenfunction, and one energy eigenfunction is lost. The $t e^{i\omega t}$ state does however still belong to the Hilbert space, with the stationary plus non-stationary solutions combined still being complete. In consequence the theory is still unitary \cite{Bender2008b}.

Further insight into the $\omega_1=\omega_2=\omega$ limit is provided by noting that all of the functions $\omega_1^2+\omega_2^2=2\omega^2$, $\omega_1^2\omega_2^2=\omega^4$, $\omega_1+\omega_2=2\omega$, $\omega_1\omega_2=\omega^2$ remain real in the limit. Consequently in the limit $\bar{H}$ as given in (\ref{2.16}) remains $PT$ symmetric, while the asymptotic behavior of the wave function $\psi_0(x,z)$ remains unchanged so that it still needs to be continued into the same Stokes wedge that is needed when $\omega_1\neq \omega_2$. Thus for $\omega_1$ and $\omega_2$  real and unequal, real and equal, or in a complex conjugate pair, the  Pais-Uhlenbeck theory is $PT$ symmetric and unitary and nowhere is there any state of negative norm.  The existence of an antilinear symmetry is thus seen to be the unifying feature, and for each of these three possible realizations one has a consistent quantum theory.

\section{Consistent Quantum Gravity Theories}
\label{S5}

There are two general approaches to constructing a consistent quantum gravity theory that are based on the above analysis, second-order plus fourth-order gravity, or pure fourth-order gravity. The typical second plus fourth approach involves a gravitational action that contains both $R$ and $R^2$ terms, and a linearization of quantum fluctuations around a flat background according to $g_{\mu\nu}=\eta_{\mu\nu}+h_{\mu\nu}$ leads to typical propagators such as the $D(k)$ one given in (\ref{2.6}). As such these theories have a potential ghost problem, but as we have seen, on constructing the relevant Hilbert space there is in fact no ghost problem after all, with these theories thus being quantum-mechanically viable.

For pure fourth-order gravity a particularly interesting case is the conformal gravity theory as it possesses a local conformal symmetry of the form $g_{\mu\nu}(x)\rightarrow e^{2\alpha(x)} g_{\mu\nu}(x)$ introduced by Weyl. For this theory  the action is uniquely of the form \cite{Mannheim2006}
\begin{eqnarray}
I_{\rm W}= -2\alpha_g\int d^4x\, (-g)^{1/2}\left[R_{\mu\kappa}R^{\mu\kappa}-\frac{1}{3} (R^{\alpha}_{\phantom{\alpha}\alpha})^2\right],
\label{5.1}
\end{eqnarray}
where $\alpha_g$ is a dimensionless  gravitational coupling constant. (For any coefficient other than $-1/3$ this action would only be globally scale invariant.) When linearized around flat spacetime, one can find a transverse-traceless gauge ($\partial_{\mu}K^{\mu\nu}=0$, $\eta^{\mu\nu}K_{\mu\nu}=0$ where $K_{\mu\nu}=h_{\mu\nu}-\tfrac{1}{4}\eta_{\mu\nu}\eta^{\alpha\beta}h_{\alpha\beta}$) in which the equation of motion is of a form 
\begin{eqnarray}
\Box^2K_{\mu\nu}=0
\label{5.2}
\end{eqnarray}
that is diagonal in the $(\mu,\nu)$ indices. In consequence, the components of the $K_{\mu\nu}$ fluctuation propagate independently, and one can associate with each one wave functions $e^{\pm ik\cdot x}$ and $te^{\pm ik\cdot x}$ (the solutions to $\Box^2K_{\mu\nu}=0$) and propagator $D(k^2)=1/k^4$. Given the non-stationary $te^{\pm ik\cdot x}$ solutions conformal gravity thus emerges as a Jordan-block theory with a non-Hermitian but $PT$-invariant (actually both $CPT$-invariant and  $C$-invariant) Hamiltonian that cannot be diagonalized. Being in the $PT$ category conformal gravity also emerges as a ghost-free quantum gravity theory, one that because of its dimensionless coupling constant  $\alpha_g$ is  renormalizable as well, all just as one would want of a quantum gravity theory. Since the theory is formulated in four spacetime dimensions (the only spacetime dimension in which $\alpha_g$ is dimensionless) it thus serves as an alternative to string theory with its as of yet undetected supersymmetry or extra dimensions.

That conformal gravity must be ghost free can be understood as follows. The Dirac action for a massless fermion coupled to a background geometry is of the form
\begin{eqnarray}
I_{\rm D}=\int d^4x(-g)^{1/2}i\bar{\psi}\gamma^{c}V^{\mu}_c(\partial_{\mu}+\Gamma_{\mu})\psi, 
\label{5.3}
\end{eqnarray}
where the $V^{\mu}_a$ are vierbeins and $\Gamma_{\mu}$ is the spin connection that enables $I_{\rm D}$ to be locally Lorentz invariant. As constructed, $\Gamma_{\mu}$ also enables $I_{\rm D}$ to be  locally conformal invariant under $V^{\mu}_a\rightarrow e^{-\alpha(x)}V^{\mu}_a(x)$, $\psi(x)\rightarrow e^{-3\alpha(x)/2}\psi(x)$, $g_{\mu\nu}(x)\rightarrow e^{2\alpha(x)} g_{\mu\nu}(x)$. We now introduce  the path integral $\int D[\psi]D[\bar{\psi}]\exp{iI_{\rm D}}=\exp(iI_{\rm EFF})$, and on performing the path integration on $\psi$ and $\bar{\psi}$ obtain an effective action whose leading term is \cite{tHooft2010a}
\begin{eqnarray}
I_{\rm EFF}&=&\int d^4x(-g)^{1/2}C\left[R_{\mu\nu}R^{\mu\nu}-\frac{1}{3}(R^{\alpha}_{\phantom{\alpha}\alpha})^2\right],
\label{5.4}
\end{eqnarray}
where $C$ is a log divergent constant. (This term would remain leading even if we were to give the fermion a mass.) We immediately recognize (\ref{5.4}) as the conformal gravity action $I_{\rm W}$ given in (\ref{5.1}), which it would have to be since we started with an $I_{\rm D}$ that was locally conformal invariant. Now $I_{\rm D}$ is linear in both $\psi$ and $\bar{\psi}$, with the path integration on $\psi$ and $\bar{\psi}$ thus corresponding to one-loop Feynman diagram in an external gravitational field. Since one cannot change the signature of a Hilbert space in perturbation theory, either $I_{\rm D}$ and $I_{\rm EFF}$ both have ghosts or neither does. But $I_{\rm D}$ is just the standard action for a Dirac fermion coupled to an external gravitational field, and it is ghost free. Hence $I_{\rm EFF}$ must be ghost free too. Conformal gravity must thus be ghost free. Now $I_{\rm D}$ is a fully standard action and a fermion path integration is also completely standard. Thus when the standard $SU(3)\times SU(2)\times SU(1)$ gauge theory ($I_{\rm D}$ with internal quantum numbers and gauge fields added in) is coupled to gravity one must generate $I_{\rm EFF}$. Conformal gravity thus must be an integral part of the standard theory and one is not free to leave it out. (Interestingly, because of the underlying conformal symmetry, in the fermion path integration on  $I_{\rm D}$ that leads to $I_{\rm EFF}$ one generates neither the non-conformal Einstein-Hilbert action or a non-conformal cosmological constant term -- with their exclusion solving both the dark matter and dark energy problems \cite{Mannheim2006,Mannheim2012,Mannheim2017}.) The consistency of the standard theory requires that it be ghost free, and thus the role of $PT$ symmetry in solving the ghost problem becomes paramount for the consistency of the standard $SU(3)\times SU(2)\times SU(1)$ theory when coupled to a gravitational field.

\end{document}